\RequirePackage{fix-cm}
\documentclass[twocolumn,epjc3,pdftex]{svjour3}

\expandafter\let\csname equation*\endcsname\relax
\expandafter\let\csname endequation*\endcsname\relax

\usepackage{amsmath,bm}

\usepackage{amssymb}
\usepackage{graphicx}
\usepackage[bb=boondox]{mathalfa}
\usepackage{hhline}
\usepackage{mathrsfs}
\usepackage{braket}
\usepackage{caption}
\usepackage[colorlinks,citecolor=blue,urlcolor=blue,linkcolor=blue]{hyperref}
\usepackage{xcolor}
\usepackage{color,soul}
\usepackage{caption}
\usepackage{subcaption}

\makeatletter
\newcommand{\mainmatter}{%
  \setcounter{footnote}{0}%
  \patchcmd{\@makefntext}{\fnsymbol}{\arabic}{}{}%
  \patchcmd{\@thefnmark}{\fnsymbol}{\arabic}{}{}%
  \def\@makefnmark{\textsuperscript{\arabic{footnote}}}%
}
\makeatother
\usepackage[T1]{fontenc}
\usepackage{microtype}
\usepackage[english]{babel}
\usepackage{tikz-cd}
\usepackage{ulem}

\smartqed  % flush right qed marks, e.g. at end of proof
\RequirePackage{graphicx}
\RequirePackage[colorlinks,citecolor=blue,urlcolor=blue,linkcolor=blue]{hyperref}
\journalname{Eur. Phys. J. C}
\usepackage[T2A]{fontenc} %потом убрать
\begin{document}

\title{Non-minimal Effective Scalar--Tensor Gravity in the Early Universe%\thanksref{t1}
}

\titlerunning{Non-minimal Effective Scalar--Tensor Gravity...}        % if too long for running head

\author{Oleg Zenin\thanksref{t1,addr1}
        \and
        Roman Stamov\thanksref{addr2}
        \and
        Sergey Kuzmin\thanksref{addr1}
        \and
        Stanislav Alexeyev\thanksref{t1,e1,addr1,addr3}
        }

\thankstext{t1}{The study was conducted under the state assignment of Lomonosov Moscow State University. The work of O. Zenin was supported by the Foundation for the Advancement of Theoretical Physics and Mathematics “BASIS”, grant No. 22-2-2-11-1.}
%Grants or other notes
\thankstext{e1}{e-mail: alexeyevso@my.msu.ru}


\institute{
Department of Quantum Theory and High Energy Physics, Physics Faculty, Lomonosov Moscow State University, Vorobievi Gory, 1/2, Moscow, 117234, Russia \label{addr1}
         \and
Department of Astrophysics and Stellar Astronomy, Physics Faculty, Lomonosov Moscow State University, Universitetskii Prospekt, 13, Moscow, 117234, Russia \label{addr2}         
         \and
Sternberg Astronomical Institute, Lomonosov Moscow State University, Univeresitesky Prospekt, 13, Moscow, 117234, Russia \label{addr3}
           }

\date{Received: date / Accepted: date}

\maketitle

\begin{abstract}
We study the consistency of several early-Universe scenarios within a framework of non-minimal effective sca\-lar--ten\-sor gravity. We show that bounce, inflation, and genesis stages are supported within the aforementioned theory. Consequently, this framework can serve as a viable model of the early Universe, where accelerated expansion is driven by the theory's own intrinsic degrees of freedom. Notably, the theory also provides two different values of the Hubble parameter, potentially explaining the different values of the Hubble constant measured from galaxy clusters and relic radiation, respectively.

%\keywords{black hole shadow \and black hole rotation \and Newman-Janis algorithm \and Horndesky theory \and bumblebee model \and Gauss-Bonnet scalar gravity}



\end{abstract}

\section{Introduction}
The existence of dark matter and dark energy motivates extensions of the General Relativity (GR) that include these components as intrinsic sources. Recent Dark Energy Spectroscopic Instrument (DESI) results have prompted renewed interest in dynamical dark energy models that deviate from the standard $\Lambda$CDM paradigm~\cite{DESI:2025fxa}. At the same time, GR successfully explains a wide range of astronomical observations. It is important to note that based on the earliest cosmological solutions~\cite{Friedman:1922kd} the Einstein field equations require the inclusion of the stress--energy tensor. The explanation of its origin often lies beyond the scope of GR, thereby motivating modifications and extensions of the theory~\cite{Alexeyev:2023,Capozziello:2011et,Berti:2015itd,Barack:2018yly,Alexeyev:2022mqb}. Therefore we consider a sca\-lar--ten\-sor theory in which the geometric sector is augmented by additional physical fields. Because such extensions typically introduce higher-order equations, we adopt a formulation where higher derivatives cancel each other. The most general theory satisfying this condition is Horndeski gravity~\cite{Horndeski:1974wa,Kobayashi:2019hrl}. After GW170817, these models (e.g., theories with a massive graviton) were strongly constrained by multimessenger observations~\cite{Ezquiaga:2017ekz,Creminelli:2017sry}. Nevertheless, Horndeski-type theories are frequently applied at redshifts ($z \neq 0.01$), corresponding to distances greater than that of the neutron star merger observed in GW170817. These theories also admit non-singular early-universe solutions called ``bounce''~\cite{Starobinsky:1980te,Ageeva:2021yik,Alexeyev:2025}. To be compatible with modern cosmological scenarios such theories must incorporate inflation~\cite{Guth:1980zm,Kobayashi:2011nu,Sato:1981qmu,Sato:1981ds,Starobinsky:1980te} or galileon genesis~\cite{Creminelli:2006xe,Creminelli:2010ba,Creminelli:2012my,Volkova:2019jlj,Libanov:2016kfc}. In the most general sca\-lar--ten\-sor frameworks, such as Horndeski or DHOST theories, the complexity and the number of equations grow substantially making them less suitable for practical astrophysical usage. Consequently, one has to look for a simpler model with the required properties. We focus on a subclass known as the ``Fab Four''. These models naturally lead to accelerated cosmic expansion~\cite{Charmousis:2011bf,Copeland:2012qf} without invoking a cosmological constant term ($\Lambda$). To describe more accurately small scales quantum field corrections are incorporated into the gravitational theory~\cite{Calmet:2015dpa,Alexeyev:2017scq}. A similar refinement has been applied to the Fab Four model~\cite{Latosh:2018xai}.

We consider a non-minimal effective sca\-lar--ten\-sor model that includes third- and fourth--order derivative terms constructed from one-loop contributions~\cite{Latosh:2020jyq}. The action for this theory is:
\begin{align}
    S & = \int \sqrt{-g}\Bigg[\left(\dfrac{2}{\kappa^2}+\alpha\phi^2\right)R + \kappa^2\beta G^{\mu\nu}\partial_{\mu}\phi\partial_{\nu}\phi \nonumber \\ 
    & - \dfrac{1}{2}g^{\mu\nu}\partial_{\mu}\phi\partial_{\nu}\phi - \dfrac{1}{3!}\lambda\phi^3-\dfrac{1}{4!}\tilde{g}\phi^4\Bigg] \, d^4x,\label{action1}
\end{align}
where $\kappa^2 = 32\pi G$ with $G$ representing Newton's gravitational constant, $\phi$ is a scalar field, $R$ is the Ricci scalar, $\alpha$ and $\beta$ are dimensionless constants, $\lambda$ is a cubic scalar coupling with mass dimension, $\tilde{g}$ is a dimensionless quartic scalar coupling, and $G_{\mu\nu}$ denotes the Einstein tensor defined as 
\begin{align*}
G_{\mu\nu} = R_{\mu\nu} - \dfrac{1}{2}g_{\mu\nu}R.    
\end{align*}
The discussed action~\eqref{action1} was derived in~\cite{Latosh:2020jyq} within the effective field theory framework. The model accounts one-loop corrections to the graviton propagator leading to the appearance of the $R^2$ and $R^{2}_{\mu\nu}$ operators. However, such corrections vanish in four-dimensional spacetime. Consequently, there are no Ostrogradsky instabilities in the theory. The addition of self-interaction terms of cubic $\phi^3$ and quartic $\phi^4$ orders allows one to study gravity in the low-energy regime.

We propose the model~\eqref{action1} as it is significantly simpler than Horndeski or DHOST theories in their general form, while preserving all their interesting key properties. Secondly, it retains the advantages of $f(R)$ gravity (including Starobinskii inflation) by providing an accelerated expansion without introducing a $\Lambda$ term. Finally, it does not require tuning the graviton propagation speed. These properties are achieved because of the existence of a non-minimal coupling $G^{\mu\nu}\partial_{\mu}\phi\partial_{\nu}\phi$ in the minimal model (see Section 2 of~\cite{Latosh:2020jyq}), known as the John interaction from the Fab Four model~\cite{Kobayashi:2019hrl}. This coupling describes the Universe accelerated expansion and notably, due to its nonlinear nature, it affects the propagation speed of gravitational waves. Using experimental data from the LIGO collaboration, the following constraint on the difference between the graviton and photon speeds was obtained:
\begin{align}
\label{BLc}
-3 \times 10^{-15} \leq 256 \beta \left(\pi G \dot{\phi}\right) \leq 7\times 10^{-16} .
\end{align}
Thus, if the condition~\eqref{BLc} is satisfied the model~\eqref{action1} appears to be consistent with modern experimental and observational results. Finally, since the John interaction is realized even with the choice of a simplified action with the minimal coupling, it can be viewed as a part of an effective theory built upon the Standard Model (where the Higgs scalar serves as a truly fundamental particle).

As a first step, we analyze a bounce-type solution~\cite{Alexeyev:2025}. A refined analysis of the Einstein equations leads to a shift in the admissible parameter range. To assess the applicability to early-Universe evolution one has to study the realization of two cosmological scenarios, namely inflation or (and) genesis. To obtain additional constraints, we exploit the so-called ``Hubble tension''~--- the discrepancy between the Hubble constant values $H_0$ measured from Cepheids~\cite{Reid:2019tiq,Riess:2021jrx} and the cosmic microwave background~\cite{Planck:2018vyg}. The two values for the Hubble parameter $H$ can be derived from the Einstein and Klein--Gordon equations, which yield a quadratic equation for $H$. As the two observed values of the Hubble constant are almost equal we assume the corresponding discriminant being close to zero~\cite{Kontou:2024tqo}.

The paper is organized as follows. In Section~2, we derive the field equations of the model~\cite{Latosh:2020jyq}. Section~3 examines a bounce-type solution. Section~4 demonstrates the viability of a genesis scenario. In Section~5, the conditions for inflation are analyzed.  Section~6 provides a further analysis of the Einstein equations in the context of the Hubble tension. Section~7 investigates the stability of the model. Section~8 presents the construction of a three-dimensional diagram illustrating the possible cosmological scenarios. Finally, Section~9 presents a discussion on the resulting constraints and our conclusions.

\section{Einstein and Klein--Gordon Equations}

The Klein--Gordon equation for the action~\eqref{action1} \cite{Kazanas:1980tx} has the following form:
\begin{eqnarray}
\label{KG}
    & - & \dfrac{1}{2!}\lambda \phi^2 -\dfrac{1}{3!}\tilde{g}\phi^3  + \Box \phi + 2\alpha \phi R \nonumber \\ 
    & - & 2\kappa^2 \beta G^{\mu\nu}\nabla_{\mu}\nabla_{\nu}\phi = 0.\label{phieq1}
\end{eqnarray}
For the further consideration we introduce the effective gravitational constant $G_{\text{eff}}(\phi)$ which depends on the scalar field as:
\begin{equation}\label{Geff}
    \dfrac{2}{\kappa^2}+\alpha\phi^2 = \dfrac{1}{16\pi G_{\text{eff}}(\phi)}.
\end{equation}

The Einstein equations for the action~\eqref{action1} are:
\begin{eqnarray}
\label{Ein}
    \mathcal{G}_{\mu\nu} & = & \dfrac{1}{16\pi G_{\text{eff}}}G_{\mu\nu} - \left(\nabla_{\mu\nu}-g_{\mu\nu}\Box\right)\alpha \phi^2  - \dfrac{1}{2}\nabla_{\mu} \phi \nabla_{\nu} \phi \nonumber \\ 
     & - & \dfrac{1}{2}g_{\mu\nu} \left( \dfrac{1}{2}(\nabla\phi)^2 + \dfrac{1}{3!}\lambda\phi^3+\dfrac{1}{4!}\tilde{g}\phi^4 \right)  \nonumber \\
    & - & \kappa^2 \beta \bigg( -\nabla_{\lambda} \nabla_{\mu} \phi \nabla^{\lambda}\nabla_{\nu}\phi + \nabla_{\mu}\nabla_{\nu}\phi \Box \phi \nonumber \\ & + &  R_{\alpha\mu\nu\beta} \nabla^{\alpha} \phi \nabla^{\beta} \phi - \dfrac{1}{2}\Bigl[ \nabla_{\mu}\phi G_{\nu\lambda} \nabla^{\lambda}\phi \nonumber \\ & + & \nabla_{\nu} \phi G_{\mu\lambda}\nabla^{\lambda} \phi\Bigr] - \dfrac{1}{2}\Bigl[ \nabla_{\mu}\phi R_{\nu\lambda} \nabla^{\lambda}\phi \nonumber \\
    &+&  \nabla_{\nu} \phi R_{\mu\lambda}\nabla^{\lambda} \phi\Bigr]  
    + \dfrac12 G_\mu{}_\nu\nabla^\lambda\phi\nabla_\lambda\phi \nonumber \\
    &+&  g_{\mu\nu} \Bigl[ R^{\alpha\beta} \nabla_{\alpha} \phi \nabla_{\beta} \phi - \dfrac{1}{2}(\Box \phi)^2 \nonumber \\
    &+& \dfrac{1}{2}(\nabla_{\alpha\beta}\phi)^2\Bigr] \bigg) = \dfrac{1}{2}T_{\mu\nu}, 
\end{eqnarray}
where $T_{\mu\nu}$ is the stress--energy tensor:
\begin{align}
T_{\mu\nu}=\dfrac{-2}{\sqrt{-g}}\dfrac{\delta(\sqrt{-g}L_m)}{\delta g^{\mu\nu}},
\end{align}
and $L_m$ is the matter Lagrangian.

\section{Bouncing Cosmological Solutions}

We start from the Friedmann-type ansatz (analogously to~\cite{Alexeyev:2012zz,Alexeyev:2020cuq}):
\begin{equation}\label{metricF}
    ds^2=dt^2-a^2(t)(dx^2+dy^2+dz^2),
\end{equation}
where the scale factor $a$ and the scalar field $\phi$ depend upon the time $t$ only. The Klein--Gordon equation~\eqref{KG} takes the form:
\begin{align}
\ddot{\phi}&=12\alpha\phi\left(\dfrac{\ddot{a}}{a}+\dfrac{\dot{a}^2}{a^2}\right)-\beta\kappa^2\left(\dfrac{\dot{a}^2}{a^2}\ddot{\phi}+2\dfrac{\dot{a}}{a}\dfrac{\ddot{a}}{a}\dot{\phi}+\dfrac{\dot{a}^3}{a^3}\dot{\phi}\right) \nonumber \\
&-3\dfrac{\dot{a}}{a}\dot{\phi}+\dfrac{1}{2}\lambda\phi^2+\dfrac16\tilde{g}\phi^3=0, \label{KGH}
\end{align}
and the Einstein field equations~\eqref{Ein}, respectively, take the form:
\begin{align}
G_{00}&=3\dfrac{\dot{a}^2}{a^2}\left(\dfrac{2}{\kappa^2}+\alpha\phi^2\right)+6\alpha\dfrac{\dot{a}}{a}\phi\dot{\phi}-\dfrac14\dot{\phi}^2 \nonumber \\
&-\dfrac92\beta\kappa^2\dfrac{\dot{a}^2}{a^2}\dot{\phi}^2+\dfrac{1}{12}\lambda\phi^3+\dfrac{1}{48}\tilde{g}\phi^4=0, \label{G00H} \\
G_{ii}&=\left(2\dfrac{\ddot{a}}{a}+\dfrac{\dot{a}^2}{a^2}\right)\left(\dfrac{2}{\kappa^2}+\alpha\phi^2\right)+2\alpha\left(\dot{\phi}^2+\phi\ddot{\phi}+2\dfrac{\dot{a}}{a}\phi\dot{\phi}\right) \nonumber \\
&+\dfrac14\dot{\phi}^2 -\beta\kappa^2\left(\dfrac{\ddot{a}}{a}\dot{\phi}^2+2\dfrac{\dot{a}}{a}\dot{\phi}\ddot{\phi}+\dfrac12\dfrac{\dot{a}^2}{a^2}\dot{\phi}^2\right) \nonumber \\
&+\dfrac{1}{12}\lambda\phi^3+\dfrac{1}{48}\tilde{g}\phi^4=0, \label{GigH} 
\end{align}

Consider the system~\eqref{KGH}, \eqref{G00H}, and~\eqref{GigH} at the bounce. Hence, the scale factor must be positive and finite: $a = \text{const} > 0$. As the bounce corresponds to a local minimum so $\dot{a} = 0$ and $\ddot{a} > 0$. Therefore the Einstein equations at the bounce point can be simplified as:
\begin{align}
&\dfrac{1}{4}\dot{\phi}^2=\dfrac{1}{12}\lambda\phi^3+\dfrac{1}{48}\tilde{g}\phi^4, \label{G000} \\
 &2\dfrac{\ddot{a}}{a}\left( \dfrac{2}{\kappa^2}+\alpha\phi^2\right) +2\alpha\dot{\phi}^2+2\alpha  \phi\ddot{\phi} + \dfrac{1}{4}\dot{\phi}^2 \nonumber \\ 
 &- \kappa^2 \beta \dfrac{\ddot{a}}{a}\dot{\phi}^2+\dfrac{1}{12}\lambda\phi^3+\dfrac{1}{48}\tilde{g}\phi^4=0, \label{G111} 
\end{align}
and the Klein--Gordon equation~\eqref{phieq1} takes the form:
\begin{equation}\label{eq19}
\ddot{a} = \dfrac{a}{12\alpha \phi} \left(\ddot{\phi} - \dfrac{1}{2}  \lambda \phi^2 - \dfrac{1}{6} \tilde{g} \phi^3 \right).
\end{equation}

We consider the scalar field being a source for the stress--energy tensor. Therefore its first derivative must vanish: $\dot{\phi} = 0$ with $\phi = \text{const}$~\cite{Alexeyev:2025}. From the equation~\eqref{G000} one obtains the new expression for the scalar field:
\begin{equation}\label{eq20}
\phi = -4\dfrac{\lambda}{\tilde{g}} .
\end{equation}
Using~\eqref{G111} and~\eqref{eq19} one obtains the expression for the second derivative:
\begin{equation}\label{eq21}
\ddot{\phi} = -\dfrac{8\lambda^3}{3\tilde{g}^2} \left(\dfrac{\dfrac{1}{\kappa^2}+8 \alpha\dfrac{\lambda^2}{\tilde{g}^2}}{{\dfrac{1}{\kappa^2}+8\alpha\dfrac{\lambda^2}{\tilde{g}^2}(1+12\alpha})}\right) \nonumber.
\end{equation}
Taking into account~\eqref{G000} and~\eqref{G111} and substituting them into~\eqref{eq19} one obtains the following inequalities:
\begin{align}
&\phi = -4\dfrac{\lambda}{\tilde{g}}, \label{final1}\\
&\ddot{\phi} = -\dfrac{8\lambda^3}{3\tilde{g}^2} \left( \dfrac{\dfrac{1}{\kappa^2}+8 \alpha\dfrac{\lambda^2}{\tilde{g}^2}}{{\dfrac{1}{\kappa^2}+8\alpha\dfrac{\lambda^2}{\tilde{g}^2}(1+12\alpha})} \right) , \label{final2}\\
&a>0, \label{final3}\\
&\ddot{a} = -\dfrac{16a\alpha\lambda^4}{3\tilde{g}^3} \left( \dfrac{1}{{\dfrac{1}{\kappa^2}+8\alpha\dfrac{\lambda^2}{\tilde{g}^2}(1+12\alpha})} \right) >0.  \label{final4}
\end{align}
The stability condition is: $\tilde{g} < 0$ (otherwise, the scalar field potential would be unbounded from below). The solution of~\eqref{final1}--\eqref{final4} imposes the following constraints on the $\alpha$ parameter:
\begin{align}\label{alp_bounce}
\alpha>0,
\end{align}
or
\begin{align}
\label{negalpha}
-\dfrac{1}{24}-\dfrac{1}{24}\sqrt{1-\dfrac{6\tilde{g}^2}{\kappa^2\lambda^2}}<\alpha<-\dfrac{1}{24}+\dfrac{1}{24}\sqrt{1-\dfrac{6\tilde{g}^2}{\kappa^2\lambda^2}}.
\end{align}
This last condition imposes a narrow range on the values of $\alpha$ leading to negative $G_{eff}$ values at late times (if a bounce exists). Therefore, one can introduce an additional constraint on the expression \eqref{Geff}:
\begin{align}
\label{alphab}
\alpha>-\dfrac{\tilde{g}^2}{8\kappa^2\lambda^2}.
\end{align}
The resulting restrictions on the parameter signs, derived from Eqs.~\eqref{final2}--\eqref{final4}, are as follows:
\begin{align}
\label{con1}
&\lambda>0, \alpha>0 \Rightarrow \phi>0, \ddot{\phi}<0, \\
\label{con2}
&\lambda<0, \alpha>0 \Rightarrow \phi<0, \ddot{\phi}>0.
\end{align}
These inequalities represent the conditions for the bounce existence.

Another important feature of this cosmological scenario is the sign change of the Hubble parameter from negative to positive at the point where the Universe passes the origin ($t = 0$) as the contraction changes to the expansion. Since $a=const>0$, it follows that $\dot{a}=0$ and $\ddot{a}>0$. Hence $H=0$, $\ddot{H}>0$. These conditions are sufficient to guarantee a change of the $H(t)$ sign as it passes through the origin.

\section{Cosmological Solutions with Genesis Stage}

We consider Taylor series of the scalar field $\phi$ near $t = 0$ applying the values of $\phi(0)$ and $\ddot{\phi}(0) \equiv \gamma$ from the bounce analysis. The first field derivative vanishes at $t = 0$ therefore:
\begin{equation}
\label{inf_5}
\phi(t) \approx \phi(0) +\dot{\phi}(0)t + \dfrac{1}{2}\ddot{\phi}(0)t^2 \equiv -4\dfrac{\lambda}{\tilde{g}}+\dfrac{1}{2}\gamma t^2. 
\end{equation}
Next we substitute the expansion~\eqref{inf_5} into~\eqref{G00H} with respect to $H$. As both values of the Hubble parameter near $t = 0$ are almost equal therefore the discriminant of~\eqref{G00H} is approximately equal to zero:
\begin{equation}
\label{HubblePar}
H(t) = \dfrac{At+Bt^3}{C-\left(A+D\right)t^2-\dfrac12Bt^4}.
\end{equation}
where
\begin{align}
\label{Par}
&A=4\alpha\dfrac{\lambda}{\tilde{g}}\gamma >0, \\
&B=-\dfrac{1}{2}\alpha\gamma^2<0 , \\
&C=\dfrac{2}{\kappa^2}+16\alpha\dfrac{\lambda^2}{\tilde{g}^2}>0, \\
&D=\dfrac32\kappa^2\beta\gamma^2.
\end{align}
Conditions~\eqref{con1} and~\eqref{con2} determine the sign of $A,B,C$ combination.

The key feature of the genesis stage is that the expansion starts from the asymptotically flat spacetime~\cite{Creminelli:2010ba}. To reproduce this scenario, the initial conditions are chosen near $t=0$, where the Hubble parameter is small and its time derivative is negligible: $H\approx 0$, $\dot{H} \approx 0$. Thus, we consider $H(t)$ Taylor series around $t = 0$ up to the third order. The equation~\eqref{HubblePar} serves as the basis for this expansion:
\begin{align}
H(t) \approx \dfrac{A}{C}t+\dfrac{A(A+D)+BC}{C^2}t^3,
\end{align}
then
\begin{align}
\dot{H}(t) \approx \dfrac{A}{C}+3\dfrac{A(A+D)+BC}{C^2}t^2.
\end{align}
Since the time value is small, we rewrite the previous expression as:
\begin{align}
\dot{H}(t) \approx \dfrac{A}{C},
\end{align}
or, using the original notations:
\begin{equation}
\dot{H}(t) \approx \dfrac{2\tilde{g}\alpha\kappa^2\gamma\lambda}{\tilde{g}^2+8\alpha\kappa^2\lambda^2}. \nonumber
\end{equation}
The condition $H \approx 0$, i.e., $\dot{H} \approx 0$ can be rewritten as:
\begin{equation}
\label{inf_8}
2\tilde{g}\alpha\kappa^2\gamma\lambda \ll \tilde{g}^2+8\alpha\kappa^2\lambda^2.   
\end{equation}
Fortunately, the last expression provides the flat spacetime phase necessary for the genesis scenario. Substituting expression~\eqref{final2} for $\gamma$ into~\eqref{inf_8}, one obtains:
\begin{equation}
\dfrac{-\dfrac{16}{3\tilde{g}^3}\alpha\kappa^2\lambda^4\left(1+8 \alpha\kappa^2\dfrac{\lambda^2}{\tilde{g}^2}\right)}{{1+8\alpha\kappa^2\dfrac{\lambda^2}{\tilde{g}^2}(1+12\alpha})} \ll 1+8\alpha\kappa^2\dfrac{\lambda^2}{\tilde{g}^2}. \nonumber     
\end{equation}
As $\left(\alpha > -\dfrac{\tilde{g}^2}{8\kappa^2\lambda^2}\right)$, i.e., $\left(1 + 8\alpha\kappa^2 \dfrac{\lambda^2}{\tilde{g}^2} > 0\right)$ without the sign changing one arrives to the following:
\begin{equation}
\label{inf_9}
\dfrac{-\dfrac{16}{3\tilde{g}^3}\alpha\kappa^2\lambda^4}{1+8\alpha\kappa^2\dfrac{\lambda^2}{\tilde{g}^2}(1+12\alpha)} \ll 1. \nonumber     
\end{equation}
Since the denominator in the left-hand side is positive then:
\begin{equation}
-\dfrac{16}{3\tilde{g}^3}\alpha\kappa^2\lambda^4 \ll 1+8\alpha\kappa^2\dfrac{\lambda^2}{\tilde{g}^2}(1+12\alpha). \nonumber   
\end{equation}
Rewrite the inequality as:
\begin{equation}
\label{gencon}
96\alpha^2+8\alpha\left(1+\dfrac23\dfrac{\lambda^2}{\tilde{g}}\right)+\dfrac{\tilde{g}^2}{\lambda^2\kappa^2} \gg 0,   
\end{equation}
Thus, we obtain a sufficient condition for the flat spacetime phase. Now it is convenient to consider the resulting inequality and derive the necessary condition. From the denominator of~\eqref{final2} it follows that the constant term is small: $0<\dfrac{\tilde{g}^2}{\lambda^2\kappa^2}<\dfrac16$, which simplifies the inequality. As a result, the following necessary condition:
\begin{equation}
\label{gencon1}
\alpha\gtrsim-\dfrac{1}{12}-\dfrac{1}{18}\dfrac{\lambda^2}{\tilde{g}}>0 \Rightarrow  \dfrac{\lambda^2}{\tilde{g}} \lesssim -\dfrac32,    
\end{equation}
appears, providing an upper bound for the combination of the discussed parameters.

The expansion phase occurs if:
\begin{equation}
A(A+D)+BC>0 \Rightarrow D>-\dfrac{BC}{A}-A, \nonumber   
\end{equation}
or, in the original notation:
\begin{equation}
\label{gencon2}
\beta>\dfrac16\dfrac{\tilde{g}}{\lambda\kappa^2\gamma}\left(\dfrac{1}{\kappa^2}-8\alpha\dfrac{\lambda^2}{\tilde{g}^2}\right) .  
\end{equation}
If $\dfrac{1}{\kappa^2}-8\alpha\dfrac{\lambda^2}{\tilde{g}^2}>0$, then this expression constrains the allowed range as $\alpha$: $0<\alpha<\dfrac18\dfrac{\tilde{g}^2}{\lambda^2\kappa^2}$. Otherwise, there is a lower bound constraint on $\alpha$: $\alpha>\dfrac18\dfrac{\tilde{g}^2}{\lambda^2\kappa^2}$. 

\section{Cosmological Solutions with Inflation Stage}

The inflationary stage is characterized by an exponential growth of the scale factor. In the discussed model the Hubble parameter satisfies the conditions $H(t) \approx \text{const}$ and $\dot{H}(t) \approx 0$. Consider the case where a bounce is followed by inflation. In this case one has to expand~\eqref{HubblePar} in a Taylor series around $t=0$ up to third order:
\begin{align}
H(t) \approx \dfrac{A}{C}t+\dfrac{A(A+D)+BC}{C^2}t^3.
\end{align}
This function has extrema at:
\begin{align}
&t_1=-\sqrt{-\dfrac{1}{3}\dfrac{AC}{A(A+D)+BC}}, \\
&t_2=\sqrt{-\dfrac{1}{3}\dfrac{AC}{A(A+D)+BC}}.
\end{align}
Since the expression under the square root is always positive the following constraints appear:
\begin{align}
\label{(A+D)+BC}
D<-\dfrac{BC}{A}-A \Longrightarrow \beta<\dfrac16\dfrac{\tilde{g}}{\lambda\gamma\kappa^2}\left(\dfrac{1}{\kappa^2}-8\alpha\dfrac{\lambda^2}{\tilde{g}^2}\right).
\end{align}

To realize an inflationary phase it is necessary that $H(t) \approx const$ in the vicinity of the point $t_2$. Therefore, $H(t)$ near this point must exhibit neither slope nor curvature which means that $\dot{H}(t_2) \approx 0$ and $\ddot{H}(t_2) \approx 0$. The first condition is satisfied automatically because $t_2$ is the root of the equation $\dot{H}(t) = 0$. Consider the second condition:
\begin{align}
\ddot{H}(t_2)=\sqrt{-\dfrac{12A[A(A+D)+BC]}{C^3}} \approx 0.
\end{align}
Taking into account~\eqref{(A+D)+BC} the expression under the square root must be non-negative so the following sufficient condition for inflation arises:
\begin{align}
C^3 \gg -12A[A(A+D)+BC], \nonumber
\end{align}
or in terms of the original parameters:
\begin{align}
\label{infcon}
\left(\dfrac{2}{\kappa^2}+16\alpha\dfrac{\lambda^2}{\tilde{g}^2}\right)^3 \gg& -48\alpha^2\dfrac{\lambda}{\tilde{g}\gamma^3}\left(8\alpha\dfrac{\lambda^2}{\tilde{g}^2}+3\kappa^2\beta\dfrac{\lambda}{\tilde{g}}\gamma-\dfrac{1}{\kappa^2}\right).
\end{align}

Additionally, it seems interesting to consider the conditions under which the genesis scenario is subsequently followed by a phase of inflation~\cite{Cai:2017,Volkova:2025uer}. So one has to expand the function~\eqref{HubblePar} in a Taylor series around $t=0$ up to fifth order using the genesis condition~\eqref{gencon}:
\begin{align}
H(t)&\approx\dfrac{A(A+D)+BC}{C^2}t^3\nonumber \\
&+\dfrac{\dfrac12ABC+(A+D)(A[A+D]+BC)}{C^3}t^5 .
\end{align}
The resulting function exhibits an extrema at:
\begin{align}
&t_0^3=0, \\
&t_{1,2}= \mp \sqrt{-\dfrac35\dfrac{C(A[A+D]+BC)}{\dfrac12ABC+(A+D)(A[A+D]+BC)}}.
\end{align}
The point $t_0$ is a triple root. It corresponds to an inflection point characterizing the genesis phase. The points $t_1$ and $t_2$ are associated with the inflationary stage for $t<0$ and $t>0$ respectively. The requirement  $H(t_2) \neq 0 $ together with the positivity of the expression under the square root  leads to the following constraints:
\begin{align}
&A(A+D)+BC>0 \nonumber\\ 
\label{betainfgen}
&\Longrightarrow \beta>\dfrac16\dfrac{\tilde{g}}{\lambda\gamma\kappa^2}\left(\dfrac{1}{\kappa^2}-8\alpha\dfrac{\lambda^2}{\tilde{g}^2}\right), \\
&\dfrac12ABC+(A+D)(A[A+D]+BC)<0.
\end{align}
The first condition is identical to~\eqref{gencon2}. The value of $t_2$ is small. This requirement together with the genesis condition \eqref{gencon} leads to:
\begin{align}
\label{geninflcon}
-\dfrac{D}{C}\gg1 \Rightarrow -D\gg C\Rightarrow D<0.
\end{align}
Thus the following constraints on $\beta$ arise:
\begin{align}
\label{geninflcon1}
\dfrac16\dfrac{\tilde{g}}{\lambda\gamma\kappa^2}\left(\dfrac{1}{\kappa^2}-8\alpha\dfrac{\lambda^2}{\tilde{g}^2}\right)<\beta<0.
\end{align}

Analogously to the previous discussion ($\dot{H}(t_2) \approx 0$ and $\ddot{H}(t_2) \approx 0$). The first condition is automatically satisfied; now consider the second:
\begin{align}
\ddot{H}(t_2)=5\dfrac{\dfrac12ABC+(A+D)(A[A+D]+BC)}{C^3}t_2 \approx 0,
\end{align}
then
\begin{align}
\sqrt{-\dfrac{27}{5}\dfrac{(A[A+D]+BC)^3}{C^3\left(\dfrac12ABC+(A+D)(A[A+D]+BC)\right)}}\approx 0.
\end{align}
Next, taking into account~\eqref{gencon} and~\eqref{geninflcon} the following constraints arise:
\begin{align}
\label{geninflcon2}
&B \gg D \Rightarrow -\alpha \gg 3\kappa^2\beta,
\end{align}

In summary, for the bounce $+$ inflation scenario realization the conditions ~\eqref{(A+D)+BC},~\eqref{infcon} must be satisfied. Preferably, the parameter $\beta$ is negative. If the genesis scenario is additionally considered, conditions~\eqref{gencon},~\eqref{geninflcon1} and ~\eqref{geninflcon2} are required.

\section{Additional Analysis of Einstein Equations}

The solutions of the Einstein equations with respect to $H$ are:
\begin{eqnarray}
\label{add_1}
H & = & \dfrac{-6\alpha\phi\dot{\phi}}{3\left(\dfrac{4}{\kappa^2}+2\alpha\phi^2-3\dot{\phi^2}\beta\kappa^2\right)}  \\ \nonumber 
&\pm & \dfrac{\sqrt{Z}}{3\left(\dfrac{4}{\kappa^2}+2\alpha\phi^2-3\dot{\phi^2}\beta\kappa^2\right)} ,  
\end{eqnarray}
where
\begin{eqnarray*}
Z & = & 36\alpha^2\phi^2\dot{\phi^2} - \left(\lambda\phi^3 + \dfrac{\tilde{g}\phi^4}{4}-9\dot{\phi^2}\right) \Biggl(\dfrac{2}{\kappa^2} \\ & + & \alpha\phi^2-\dfrac{3\beta\kappa^2\dot{\phi^2}}{2}\Biggr).    
\end{eqnarray*}

If $\beta$ is negative (therefore the denominator~\eqref{add_1} positive) the existence of two positive Hubble parameter values requires the following:
\begin{align}
\label{add_2}
& \left(\lambda\phi^3 + \dfrac{\tilde{g}\phi^4}{4}-9\dot{\phi^2}\right)\Biggl(\dfrac{2}{\kappa^2}+\alpha\phi^2-\dfrac{3\beta\kappa^2\dot{\phi^2}}{2}\Biggr)>0,  \\ 
%& \Big(\dfrac{\lambda\phi}{\dot{\phi^2}} + \dfrac{\tilde{g}\phi^2}{4\dot{\phi^2}}-\dfrac{9}{\phi^2}\Big)\Big(\dfrac{2}{\kappa^2}+\alpha\phi^2-\dfrac{3\beta\kappa^2\dot{\phi^2}}{2}\Big) < 36\alpha^2, \\ 
\label{add_3}
&\dfrac{2}{\kappa^2}+\alpha\phi^2-\dfrac{3\beta\kappa^2\dot{\phi^2}}{2} > 0, \\
\label{add_4}
&\lambda\phi^3 + \dfrac{\tilde{g}\phi^4}{4}-9\dot{\phi^2} > 0.
\end{align}
Since $\alpha>0$ from~\eqref{Geff} it follows that:
\begin{equation}
\label{add_55}
\phi^2 > -\dfrac{2}{\alpha\kappa^2} .     
\end{equation}
Next, from \eqref{add_4}:
\begin{align}
\label{add_5}
\dot{\phi}^2<\dfrac19\phi^3\left(\lambda+\dfrac14\tilde{g}\phi\right).
\end{align}
Assuming the parameter combination~\eqref{con1}, the condition that the right-hand side of inequality~\eqref{add_5} cannot be negative leads to:
\begin{align}
\label{add_6}
\phi>-4\dfrac{\lambda}{\tilde{g}} \Rightarrow \phi^2>16\dfrac{\lambda^2}{\tilde{g}^2},
\end{align}
which gives a lower bound for the scalar field for the parameter combination \eqref{con2}:
\begin{align}
\label{add_7}
\phi<-4\dfrac{\lambda}{\tilde{g}} \Rightarrow\phi^2<16\dfrac{\lambda^2}{\tilde{g}^2}.
\end{align}

Combining the upper and lower limits on the scalar field from \eqref{add_7} and~\eqref{add_55} into a single chain one obtains:
\begin{equation}
 -\dfrac{2}{\alpha\kappa^2}<  16\dfrac{\lambda^2}{\tilde{g}^2}. \nonumber  
\end{equation}
Thus, we obtain the following constraint, corresponding to~\eqref{alphab}:
\begin{equation}
\label{add_8}
\alpha > -\dfrac{\tilde{g}^2}{8\kappa^2\lambda^2}.    
\end{equation}
In the case $\beta > 0$ the results are the same.

\section{Stability Analysis}

To check the stability near $t = 0$ consider small perturbations around a stationary solution. So the scalar field and the Hubble parameter represent the sum of their background values and small perturbations:
\begin{align}
&\phi(t)=\phi_0+\delta\phi(t), &&H(t)=H_0+\delta H(t),\nonumber \\
&\dot{\phi}=\delta\dot{\phi}, &&\dot{H}=\delta\dot{H}, \nonumber \\
&\ddot{\phi}=\delta\ddot{\phi}, &&\ddot{H}=\delta\ddot{H}.
\end{align}
Expanding the field equations~\eqref{KGH}--\eqref{GigH} to first perturbation order the Klein--Gordon equation takes the form:
\begin{align}
\label{StKG}
\delta\ddot{\phi}=12\alpha\phi_0\delta\dot{H}+\left(\lambda\phi_0+\dfrac12\tilde{g}\phi_0^2\right)\delta\phi.
\end{align}
The time component of the Einstein tensor becomes:
\begin{align}
\label{StG00}
\left(\dfrac14\lambda\phi_0^2+\dfrac{1}{12}\tilde{g}\phi_0^3\right)\delta\phi=0,
\end{align}
and the space component is:
\begin{eqnarray}
& 2\delta\dot{H}\left(\dfrac{2}{\kappa^2}+\alpha\phi_0^2\right)+2\alpha\phi_0\delta\ddot{\phi} \nonumber \\  
&+\left(\dfrac14\lambda\phi_0^2+\dfrac{1}{12}\tilde{g}\phi_0^3\right)\delta\phi=0. \label{StGij}
\end{eqnarray}
Combining~\eqref{StKG}--\eqref{StGij} we obtain the equation for a harmonic oscillator:
\begin{align}
\delta\ddot{\phi}+\left[\dfrac{\left(\lambda\phi_0 -\dfrac12\tilde{g}\phi_0^2\right)\left(\dfrac{2}{\kappa^2}+\alpha\phi_0^2\right)}{\dfrac{2}{\kappa^2}+\alpha\phi_0^2+12\alpha^2\phi_0^2}\right]\delta\phi=0 ,
\end{align}
where the coefficient before $\delta\phi$ must be positive in order to describe stable oscillations rather than exponential growth:
\begin{align}
\dfrac{\left(\lambda\phi_0 -\dfrac12\tilde{g}\phi_0^2\right)\left(\dfrac{2}{\kappa^2}+\alpha\phi_0^2\right)}{\dfrac{2}{\kappa^2}+\alpha\phi_0^2+12\alpha^2\phi_0^2} > 0.
\end{align}
Substituting the expression~\eqref{final1} into this inequality leads to:
\begin{align}
-\dfrac{4\dfrac{\lambda^2}{\tilde{g}}\left(\dfrac{1}{\kappa^2}+4\alpha\dfrac{\lambda^2}{\tilde{g}^2}\right)}{\dfrac{1}{\kappa^2}+8\alpha\dfrac{\lambda^2}{\tilde{g}^2}(1+12\alpha)} > 0.
\end{align}
The obtained condition is fully consistent with~\eqref{con1} and \eqref{con2}. However, considering~\eqref{negalpha} leads to tighter restrictions on the $\alpha$ possible values:
\begin{align}
-\dfrac{1}{24}-\dfrac{1}{24}\sqrt{1-\dfrac{6\tilde{g}^2}{\kappa^2\lambda^2}}<\alpha<-\dfrac{1}{4}\dfrac{\tilde{g}^2}{\lambda^2\kappa^2}.
\end{align}
Therefore, the model is stable in the vicinity of $t = 0$ and the stability conditions coincide with the bounce conditions.

\section{Intersection of cosmological scenarios}

\begin{figure*}[t]
\centering
\includegraphics[width=\textwidth]{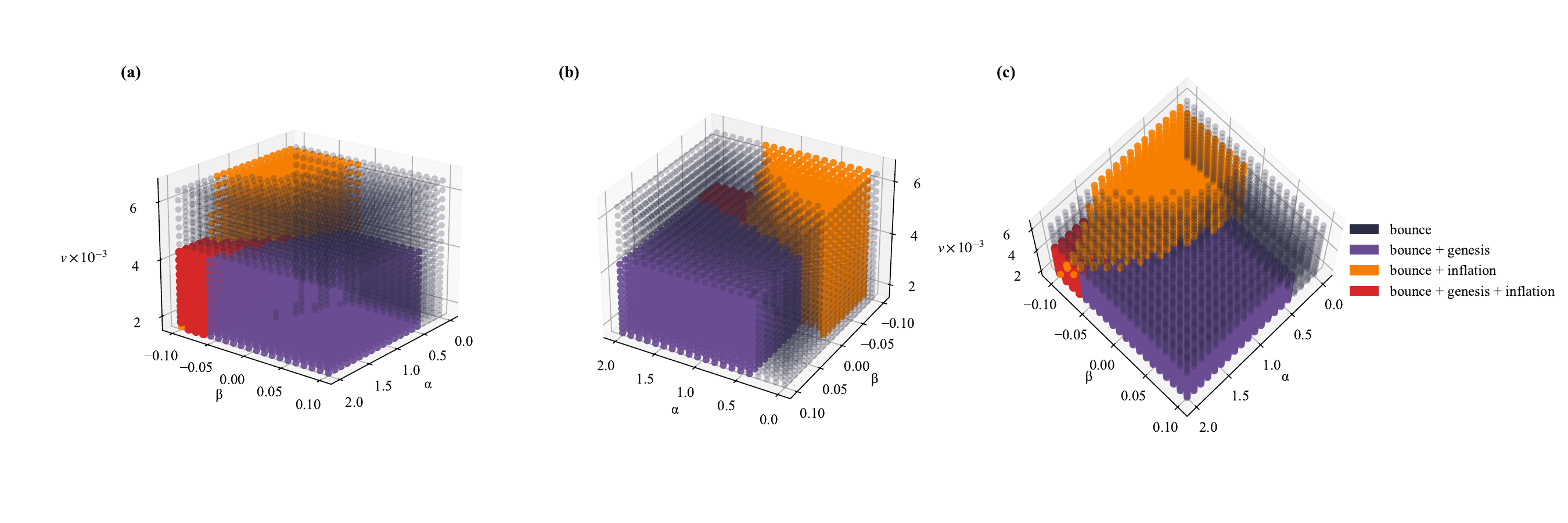}
\caption{
Three-dimensional parameter space $(\alpha,\beta,v)$
for $\tilde{g}\in[-0.8,-0.4]$, where $v\equiv\dfrac{\tilde{g}^2}{\lambda^2\kappa^2}$. Other parameters are $\lambda=1$ and $\kappa^2=32\pi$.
}
\label{fig:3D}
\end{figure*}

\begin{figure*}[t]
\centering
\includegraphics[width=\textwidth]{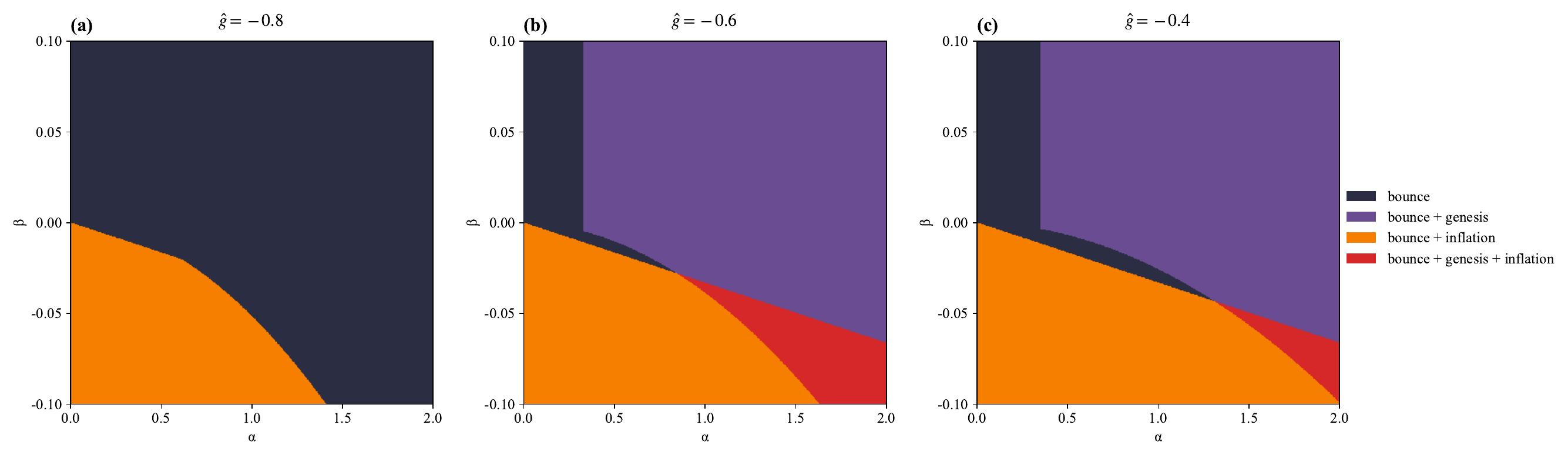}
\caption{
Phase space classification in the $(\alpha,\beta)$ plane 
for three values of $\tilde{g}$:
(a) $\tilde{g}=-0.8$, 
(b) $\tilde{g}=-0.6$, 
(c) $\tilde{g}=-0.4$, where $v\equiv\dfrac{\tilde{g}^2}{\lambda^2\kappa^2}$.
Other parameters are $\lambda=1$ and $\kappa^2=32\pi$.
}
\label{fig:slices}
\end{figure*}

To illustrate the results, we perform a qualitative visualization of the constraints on the three cosmological scenarios and construct a diagram in the parameter space $(\alpha,\beta,v\equiv \dfrac{\tilde{g}^2}{\lambda^2\kappa^2})$, where $\kappa^2=32\pi$ ($c=G=1$ in Planck units). For the bounce scenario we use the constraints~\eqref{alp_bounce} (equation~\eqref{negalpha} imposes overly restrictive bounds for $\alpha$ and therefore is not used to plot the diagram) and~\eqref{alphab}. Equations~\eqref{con1} and~\eqref{con2} are also not used in the construction of the diagram. In what follows, we use Eq.~\eqref{final2} as the expression for $\gamma$ appearing in the genesis and inflation constraints. Depending on the sign combinations of the parameters in~\eqref{con1} and~\eqref{con2}, we obtain a system of inequalities~\eqref{final1}, \eqref{final2}, and~\eqref{final4}. It has a solution, if $0<v<\dfrac{1}{6}$, which provides us with constraints on $v$. For the genesis scenario we use constraints~\eqref{gencon}, \eqref{gencon1} and~\eqref{gencon2}. The last condition also implies $\alpha>0$ while excluding the boundary value $\alpha=\dfrac{v}{8}$. However, inequality~\eqref{gencon} takes the form $F\gg0$, which is not a quantitative estimate but a qualitative one. For visualization purposes, we therefore replace the condition $F\gg0$ by the quantitative criterion $F\ge10^p$. We choose $p=1$ to enhance the visibility of the region where all three cosmological scenarios intersect. Inflation is possible if inequalities~\eqref{geninflcon1} and~\eqref{geninflcon2} are satisfied in the equivalent form $|\alpha|\ll 3\kappa^2 \beta$, which can also be written as $|\alpha|\leq 10^p\times3\kappa^2 \beta$ with $p=1$. To illustrate these constraints, we construct a three-dimensional plot in various projections (Fig.~\ref{fig:3D}). The diagram shows that, for the considered values of the additional parameters of the theory, the model admits a bounce, a bounce accompanied by genesis, a bounce accompanied by inflation, or the realization of all three scenarios as a sequence. For illustrative purposes, cross-sections of the diagram were generated corresponding to three distinct values of $\tilde{g}$ for identical values of $\lambda$ (Fig.~\ref{fig:slices}). Note that at $\tilde{g}=-0.3$ bounce or bounce and inflation can occur and that at two other values of $\tilde{g}$ all three cosmological scenarios are realized. Certain regions of the $(\alpha,\beta)$ parameter space allow the sequential realization of bounce, genesis, and inflation.

\section{Discussion and Conclusions}

We consider a non-minimal sca\-lar--ten\-sor model of gravity with third- and fourth--order field terms arising from the summation of one-loop interactions~\cite{Latosh:2020jyq}. Being a particular case of Horndeski theory, this model belongs to the Fab Four class where cosmic expansion is realized without the introduction of a cosmological constant. At the same time, due to one-loop contributions the propagation speed of gravitational waves remains consistent with the experimental constraints. Therefore, this model appears to be a good candidate to extend GR without additional screening mechanisms. Next, although it constitutes a restricted subset of the general Horndeski theory, it represents a more suitable candidate for an extended theory of gravity possessing intrinsic mechanisms for inflation, accelerated expansion and other shortcomings of GR both in early Universe and for the current stage.

In this work we concentrated on the predictions for the early Universe. Particularly, it is shown that the discussed theory allows for the realization of a bounce with genesis or a bounce with inflation. First of all we refine the previously obtained conditions for a bounce existence~\cite{Alexeyev:2025}, showing that it is also realized for $\lambda < 0$, $\tilde{g} < 0$, $\alpha > 0$ or $\lambda > 0$, $\tilde{g} < 0$, $\alpha > 0$. The bounce is also realized for $\alpha<0$. However, in this case, the range of the allowed $\alpha$ values is narrow and requires fine-tuning. The new limit coincides with the conditions obtained earlier~\cite{Sushkov:2023aya}\footnote{The bounce-type solution is realized in the case $\Lambda = 0$; the new results partially overlap with obtained for a simpler version of the model discussed earlier.}. Moreover, with respect the the data on the Hubble Tension a time interval was determined during which the function $H(t)$ satisfies the bounce scenario. We show that a genesis scenario can be realized with the same constraints as a bounce.

Furthermore, sufficient conditions for the realization of both genesis and inflation scenarios were obtained. They do not contradict the bounce ones. The constraints for inflation are more stringent: the necessary condition for inflation is~\eqref{(A+D)+BC}, then the parameter $\beta$ is negative (otherwise, a requirement for fine-tuning of the $\alpha$ parameter appears). This condition is already satisfied for the bounce~\cite{Sushkov:2023aya}. Thus, our results in this regard agree with those previously obtained for a simpler model also in this part. Restrictions on the scalar field value $\phi$ and $\alpha$ are derived in a manner consistent with the initial method of analysis. Finally, the stability analysis of the equations in the vicinity of $t = 0$ shows that the model remains stable for $\lambda < 0$, $\tilde{g} < 0$, $\alpha > 0$ or $\lambda > 0$, $\tilde{g} < 0$, $\alpha > 0$.

Thus, within the framework of the discussed sca\-lar--ten\-sor gravity model, it is possible to realize all three popular cosmological scenarios (bounce + inflation or bounce + genesis or bounce + genesis + inflation (when the role of genesis is to put the potential to a position for a slow roll start~\cite{Volkova:2025uer})). The conditions for the genesis and the bounce phases are identical. This means that whenever a bounce is possible it will always be followed by an inflationary or genesis phase (if $\beta > 0$ then the inflation is forbidden and genesis is automatically realized). Essentially, this model~--- with a relatively simple structure compared to other sca\-lar--ten\-sor gravity theories~--- resolves both the initial singularity problem while providing a natural mechanism for the emergence of either inflation or genesis. Therefore, this theory seems to be a promising candidate for further explanation of cosmological and astrophysical phenomena, potentially bringing us closer to the creation of a general extended gravity theory providing explanations to unresolved GR problems and being rather simple to be used in astrophysics and cosmology.

\section{Acknowledgements}

The study was conducted under the state assignment of Lomonosov Moscow State University.

The work of O. Zenin was supported by the Foundation for the Advancement of Theoretical Physics and Mathematics “BASIS”, grant No. 22-2-2-11-1.

The authors would like to thank Egor Pluzhnikov for useful discussions on the subject of the paper.

\bibliographystyle{spphys}
\bibliography{References}
%\bibliographystyle{iopart-num}
%\textbf{% BibTeX users please use one of
%\bibliography{References.bib}   }

\end{document}